# Ewald sphere construction for structural colors

**LUKAS MAIWALD,**[1,*] **SLAWA LANG,**[1] **DIRK JALAS,**[1] **HAGEN RENNER,**[1] **ALEXANDER YU. PETROV**[1,2] **AND MANFRED EICH**[1,3]

[1]*Institute of Optical and Electronic Materials, Hamburg University of Technology, Eissendorfer Strasse 38, 21073 Hamburg, Germany*
[2]*ITMO University, 49 Kronverkskii Ave., 197101, St. Petersburg, Russia*
[3]*Institute of Materials Research, Helmholtz-Zentrum Geesthacht, Max-Planck-Strasse 1, Geesthacht, D-21502, Germany*
*\*lukas.maiwald@tuhh.de*

**Abstract:** Disordered structures producing a non-iridescent color impression have been shown to feature a spherically shaped Fourier transform of their refractive-index distribution. We determine the direction and efficiency of scattering from thin films made from such structures with the help of the Ewald sphere construction which follows from first-order scattering approximation. This way we present a simple geometrical argument why these structures are well suited for creating short wavelength colors like blue but are hindered from producing long wavelength colors like red. We also numerically synthesize a model structure dedicated to produce a sharp spherical shell in reciprocal space. The reflectivity of this structure as predicted by the first-order approximation is compared to direct electromagnetic simulations. The results indicate the Ewald sphere construction to constitute a simple geometrical tool that can be used to describe and to explain important spectral and directional features of the reflectivity. It is shown that total internal reflection in the film in combination with directed scattering can be used to obtain long wavelength structural colors.

## 1. Introduction

The term structural color refers to colors which are not a result of selective absorption but of constructive and destructive interference generated by structural features. Structural coloration is found widely in nature [1-4], especially in insects [5-12] and bird feathers [13-24], but also in animal skin [25-27] and plants [28]. Many examples from nature show strong iridescence (i.e., their perceived color strongly depends on the observation angle). These iridescent structural colors are usually produced by nanostructures ordered on the same length-scale as the wavelengths of visible light (e.g. [5,12]). There are, however, also examples for structural colors which show only very weak or no iridescence under daylight illumination. It has been shown that the structures responsible for this effect often exhibit an isotropic shape of their Fourier transform whereas ordered structures are anisotropic in *k*-space. The isotropic Fourier transforms can for example be broadband, corresponding to white structural color [7], or they can be spherically shell shaped, which leads to wavelength selectivity in the reflectivity spectrum. This shell shape indicates that there is actually a kind of order in such structures. This may be caused by disordered structures with short-range order [13–20,24,25] or by random arrangements of small ordered substructures, i.e. polycrystals [6,8,11]. Non-iridescent colors produced by such structural features are mostly short- and intermediate-wavelength colors like violet, blue and green. Examples for longer wavelength colors solely produced by structure are very rare [25,29]. While such colors may comprise influences from structural features they also usually contain selective absorbers influencing their color appearance [4,13-15,23-25].

Non-iridescent structural colors may be highly interesting for decoration purposes since they allow for producing diverse colors with one single material system by simple scaling of geometry. Since, in contrast to regular pigments, the required features of such a material system comprise no complex absorption properties, simple and environmentally friendly material systems could be envisaged. As structural colors can be realized from almost non-absorbing dielectric materials, photo-bleaching can easily be avoided, making them very long-living in comparison to conventional organic pigments. In case of ceramic dielectrics, a high temperature stability can be envisaged additionally which can help to avoid the usage of toxic pigments in order to achieve coloration of ceramic products.

Current research focuses mainly on photonic glasses, which are disordered assemblies of monodisperse, i.e. equally sized spherical particles [30–32]. The monodispersity of the touching particles leads to short range order in this case. To increase the color saturation of the photonic-glass films, the introduction of broadband absorbers [33–39], optimization of film thickness and background [40,41] and the usage of more complex particles, e.g., core-shell particles [37,42,43] are discussed. However, the gamut of structural coloration is limited since so far there is no procedure known to produce highly saturated long wavelength colors like red or yellow [29].

In 1998 it was found that there are non-iridescent structural colors in nature that are not resulting from incoherent scattering. Instead, there is a short-range order leading to coherent scattering [18]. Since then, different first-order scattering approaches have been discussed [16,29,44]. We think, however, that the first-order approximation [45] was not used up to its full potential. The Ewald sphere construction, well known in X-ray scattering [46,47] and also sometimes used in optics, e.g. in holography [48,49], photonic quasicrystals [50], light extraction from LEDs [51] or nonlinear optics [52] was not fully applied to and described in detail for structural colors.

We show that by rigorous first-order derivation a formulation can be found that allows making quantitative predictions of angle dependent scattering from a volume of disordered medium. Utilizing this approach, we explain the generation of short-wavelength structural colors and point out the difficulties at long wavelengths. We focus on non-iridescent colors similar to those found in nature and propose a model structure dedicated to producing a spherical shell-shaped absolute square of the Fourier transform. For the sake of simplicity this spherical shell in $k$-space will be called $k$-shell throughout this paper. We use brute-force finite integration technique (FIT) to assess the applicability of first-order scattering to structural coloration. Based on the obtained frequency-dependent scattering directions and the interplay with total internal reflection at the film boundary a mechanism for long wavelength color generation is proposed.

## 2. Theory

The Ewald sphere construction is based on the first-order Born approximation [45]. It is a well-known and commonly used tool in X-ray diffraction and crystallography. While its application to structural colors was mentioned before [44] and similar approaches based on the spatial Fourier transform of the structure have been used (e.g. [6,53]) there has not been a detailed analysis of its implications for the explanation of structural coloration so far. Also, especially when talking about flat colored films, Fresnel reflection at the film surface may influence the color impression significantly. With the usual presentation of the first-order approximation [45] it is impossible to take this into consideration as the Fresnel reflection is a strong effect. For these reasons we use the quantitative approach recently derived in reference [54]. Note that we will discuss refractive-index distributions here while in the derivation in [54] permittivities are discussed. An arbitrary distribution of the relative permittivity in space may be described as $\varepsilon(\mathbf{r}) = \bar{\varepsilon} + \hat{\varepsilon} f(\mathbf{r})$, where $\bar{\varepsilon}$ denotes the average permittivity and $\hat{\varepsilon} f(\mathbf{r})$ is a spatially varying perturbation as a function of the position $\mathbf{r}$ with amplitude $\hat{\varepsilon}$. Assuming now $\hat{\varepsilon}$ to be small, we can alternatively describe the structure by the refractive index $n(\mathbf{r}) = \bar{n} + \hat{n}f(\mathbf{r})$, where $\bar{n} = \bar{\varepsilon}^{1/2}$ denotes the average refractive index and $\hat{n} = \hat{\varepsilon}/(2\bar{\varepsilon}^{1/2})$.

We will identify the scattering directions predicted by the first-order approximation using the Ewald sphere construction. It can be exemplified schematically with a Bragg mirror with a weak refractive index contrast, as illustrated in Fig. 1. A simple Bragg mirror is comprised of a sinusoidal variation of the refractive index. The Fourier transform of the sinusoidal refractive-index distribution in real space (top panel) produces two symmetric spots in $k$-space (bottom panel) at $\pm 2\pi/a$ on the axis defined by the direction of the refractive-index modulation, where $a$ is the lattice constant. Also shown are two waves with different wavelengths impinging on the Bragg mirror under the same angle. In order to draw the Ewald sphere the $k$-vector of an incident wave is drawn in reciprocal space such that it points at the origin of the reciprocal space. The Ewald sphere is then the sphere around the starting point of the $k$-vector with its radius corresponding to the length of the $k$-vector. It thereby represents all potential directions of elastic scattering. Then, the directions from the center of the Ewald sphere to where it intersects the Fourier transform represent the actual directions of scattering. In the left case the Ewald sphere intersects with the Fourier transform of the mirror, marked by the upper spot in reciprocal space (bottom panel), which leads to strong scattering. The direction of scattering is defined by the connection of the Ewald sphere's center to the intersection region which results in the only existing scattered wave vector $\mathbf{k}_1$. In the right case no light is scattered since the Ewald sphere only occupies regions in reciprocal space where the Fourier transform is zero. The findings are consistent with Bragg's law [45].

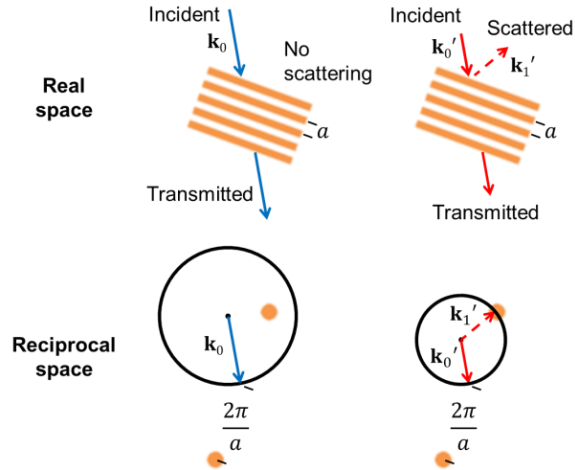

Fig. 1. Ewald sphere application for the example of a Bragg mirror with a low refractive index contrast and incident waves with two different wavelengths indicated by the arrow color. The incident wave vectors $\mathbf{k}_0$ and $\mathbf{k}_0$' are indicated by the solid arrow, the scattered wave vector $\mathbf{k}_1$' corresponds to the dashed arrow. The left short-wavelength case shows a $k$-vector for which Bragg reflection occurs while the right long-wavelength case is off Bragg resonance. In the reciprocal space, the Ewald sphere overlaps with the non-zero Fourier-transform spots in which case light is scattered back in the corresponding direction (left) or it does not overlap with the spots in which case nothing is scattered (right) and light is simply transmitted.

In the following we focus on structures which, similarly to structures found in nature, have the shape of a spherical shell ($k$-shell) of their permittivity distribution in the $k$-space. As already stated in the introduction, this shape indicates that there is a characteristic length $a$ in the structure which is found along all directions and connected to the radius $K$ of the $k$-shell by $K = 2\pi/a$. The Ewald sphere construction can be used to predict the wavelength and angle dependent scattering of such a structure under plane wave illumination. For short wavelengths, light is scattered forward (i.e. into the lower half space) such that the wave vectors of scattered light lie on the surface of a cone (Fig. 2(a)). The opening angle of this cone increases when longer wavelengths, thus shorter wave vectors, are considered until the situation in Fig. 2(b) is reached, where light is perpendicularly scattered. For even longer wavelengths the Ewald sphere shrinks further leading to scattered wave vectors lying on the surface of a cone that is now directed upwards (i.e. backscattering). At the point where the length of the incident wave vector and therefore the radius of the Ewald sphere equals half the radius of the $k$-shell backscattering into a small solid angle should be observed. The scattering for this situation is shown in Fig. 2(c); the scattering cone is filled with wave vectors due to the finite thickness of the $k$-shell. For even longer wavelengths the Ewald sphere eventually does not intersect with the Fourier transform anymore and incident light is just transmitted. It should be noted that even structural colors that are found to be non-iridescent under diffuse illumination show some iridescence under directional illumination [16]. Hence it is not surprising that with the Ewald sphere construction a change in scattering directions depending on the wavelength is predicted.

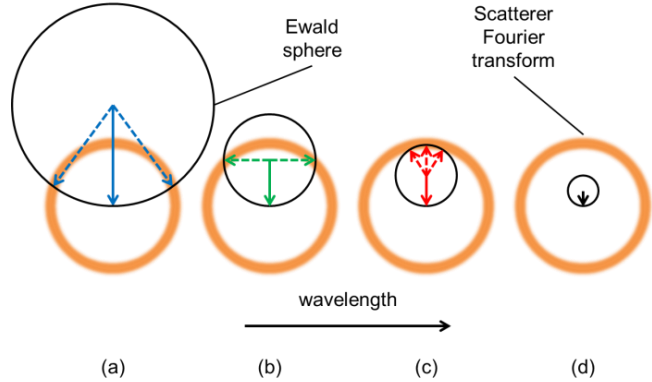

Fig. 2. 2D cut through the Ewald sphere construction for the scattering from a structure with a spherical shell shaped Fourier transform. (a) Short wavelength, leading to forward scattering, (b) intermediate wavelength, leading to perpendicular scattering, (c) long wavelength, leading to backscattering and (d) very long wavelength, leading to no scattering. Solid arrows indicate the wave vectors of the incident waves; dashed arrows indicate wave vectors of scattered waves.

From this it is already possible to draw conclusions on the reflectivity spectrum of the scattering structure. There must be a reflectivity cut-off towards long wavelengths at the point where the Ewald sphere stops intersecting with the $k$-shell. We aim at such a cut-off in order to produce a color different from white. If this cut-off is in the visible close to the UV-transition pure blue colors result, which fits to the observation that natural structures having a shell-shaped Fourier transform indeed often exhibit pure blue colors [13]. At the same time, this fact explains why pure colors featuring longer-wavelength reflection cannot easily be observed, since the part between the cut-off and the UV limit of the spectral sensitivity of the human eye contains other visible wavelengths which mix and thus impede a pure color impression.

However, when scattering occurs in a plano-parallel slab of structured material, the situation is more complicated. Two important effects should then be taken into account (see Fig. 3). First, with increasing opening angle of the scattering cone for decreasing wavelengths the backscattered light will at some point be totally internally reflected at the boundary to air. This brings about a wavelength-selective trapping of light and provides an additional cut-off towards short wavelengths. Such a second cut-off would be required to prevent shorter wavelengths from being reflected and added to the color mix, thus compromising the purity of red color, for example. Second, depending on the thickness of the scattering volume as well as on the direction of scattered light, the light path can become long enough to make an additional scattering event highly probable [17]. Such doubly scattered light of shorter wavelengths may then be redirected in such a way that it is not totally internally reflected anymore, thus can be coupled out and will add to the color mix seen by the observer. Assume, e.g., that for the borderline case in Fig. 2(b) there is a secondary scattering event. This can be modeled by just another Ewald sphere construction where the incident wave is now the original scattered one travelling parallel to the structure surface. This results in the same relative scattering directions, perpendicular to the incidence direction, meaning 50 percent of the double-scattered light is scattered to the upper half space. This light is subject to Fresnel reflection and partly to total internal reflection again, but a considerable part may now be coupled out. The outcoupled, double-scattered light therefore adds to the reflectivity of the whole structure thus to the color mix. While the total internal reflection within thin films helps to create a second cut-off at shorter wavelengths, the higher-order scattering makes the generation of long wavelength non-iridescent structural colors such as yellow and red much more difficult and may indicate why there are hardly any examples of those in nature. The multiple scattering can be suppressed by broadband absorption in the disordered media, as light scattered multiple times accumulates large optical paths. This explains why broadband absorption can help to improve structural color at longer wavelengths [38,39].

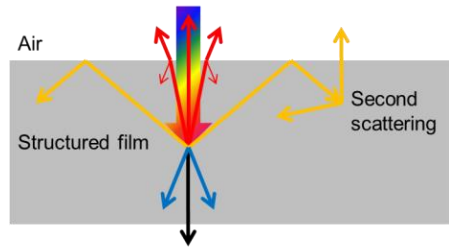

Fig. 3. Schematic 2D representation of the scattering in a film as a function of wavelength. The film is structured in such a way that its Fourier transform is a spherical shell (not shown here, see Fig. 2). Short wavelengths (shown as blue) are scattered in a forward direction. Intermediate wavelengths (orange) are backscattered in a large angle. Due to their scattering angle they may be reflected at the film surface. Also, their probability for a second scattering event is high due to a long propagation path inside the structure. Long wavelengths (red) are directly scattered back and very long wavelengths (infrared, shown as black) are not scattered at all and will just be transmitted into the downward direction.

The Ewald sphere construction can be used to make simple geometrical calculations of the scattering angles at certain wavelengths. For example, the long-wavelength edge of the direct backscattering peak corresponds to the wavelength where the length $k$ of the $k$-vector equals half the radius $K$ of the $k$-shell, $k = K/2$. The radii of the Ewald sphere and the $k$-shell are $k = 2\pi\bar{n}/\lambda_0$ and $K = 2\pi/a$, respectively, with $a$ being the effective spatial period characteristical of the scattering structure and $\lambda_0 = 2\pi c_0/\omega$ being the free space wavelength. It follows that the back reflection sets in at $\lambda_0/\bar{n} = 2a$. Another simple example is the scattering at 90 degrees, which corresponds to the case where the length of the incident wave's $k$-vector is by a factor $2^{1/2}$ smaller than the radius of the $k$-shell. It will be shown later that similar considerations can be used to explain major parts of the reflectivity spectrum of the investigated structures.

## 3. Model Structure

To demonstrate the principles underlying structural coloration we used a model structure dedicated to producing a thin *k*-shell. For this, a large number of sinusoidal refractive-index gratings of equal amplitudes $n_0$ and equal periods *a* but with random directions $\mathbf{\gamma}_i / |\mathbf{\gamma}_i|$ and phases $\phi_i$ are superimposed:

$$\hat{n}f(\mathbf{r}) = n_0 \sum_{i=1}^{N} \sin(\mathbf{\gamma}_i \cdot \mathbf{r} + \phi_i) \qquad (1)$$

The numerically resulting distribution of refractive-index values turns out to have approximately the shape of a Gaussian distribution. Thus, we characterize the structure by the standard deviation $\sigma$ of its refractive-index distribution. In case of infinitely extending gratings, each summand in Eq. (1) corresponds to two Dirac delta functions in *k*-space at $\pm \mathbf{\gamma}_i$ with $|\mathbf{\gamma}_i| = 2\pi / a = K$. Thus, we generate not a continuous shell in reciprocal space but rather a set of discrete points at equal distance to the origin of *k*-space. For finite, cuboidal segments of such a structure the Dirac functions are convoluted with sinc-functions in all spatial dimensions, leading to a broadening of the spots. With increasing number of gratings, the Fourier transform converges to a spherical shell in *k*-space. Figure 4 shows an example of a resulting structure in real space. A cut through the square of the Fourier transform of this structure at $k_z = 0$ and the average squared Fourier transform as a function of distance to the origin are shown in Fig. 5.

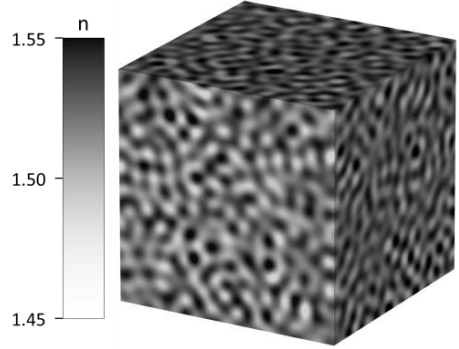

Fig. 4. 3×3×3 µm³ excerpt from the model structure made by overlap of 10,000 sinusoidal gratings with 220 nm period. The refractive-index distribution in the structure is scaled such that its standard deviation is $\sigma = 0.025$. The shown refractive-index range is limited to 1.45 to 1.55 in order to make the structure more distinguishable, the full refractive-index range considered in the calculation is from 1.40 to 1.60.

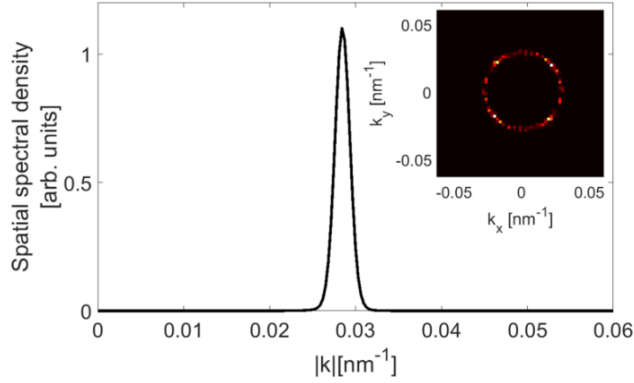

Fig. 5. Inset: Cut through the spatial spectral density of the Fourier transform of the 3×3×3 μm³ refractive index perturbation shown in Fig. 4. The plane $k_z = 0$ is shown. Main panel: Average of the spatial spectral density as a function of the distance to the origin of the reciprocal space.

The model structure we used for simulation was produced as described in Eq. (1) with parameters of $n_0 = 3.53 \times 10^{-4}$, $N = 10{,}000$ and $a = 220$ nm. An average refractive index of $\bar{n} = 1.5$ was chosen which is in the range of typical organic compounds like photosensitive polymers. The resulting distribution of the refractive index has a standard deviation of $\sigma = 0.025$; a low refractive-index contrast is needed such that the first-order Born approximation quantitatively resembles the total scattered field. Extreme points beyond $\bar{n} \pm 4\sigma$ in the refractive-index distribution have been reset to just this value of $\bar{n} \pm 4\sigma$ in order to exclude refractive indices below one or at unreasonably high values. With the assumption of a Gaussian distribution this range contains more than 99.99 percent of all refractive-index values so only very few points are affected by this.

The structure corresponds to a spherical shell in the Fourier spectrum with its maximum at $|k| = 0.0284 \text{ nm}^{-1} \approx K$ (cf. Fig. 5). We chose these parameters to get a long-wavelength reflectivity edge at a free-space wavelength $\lambda_0 = 2\bar{n}a = 660$ nm which is in the red. By simply scaling the geometrical parameters of the structure the whole spectrum could also be shifted to arbitrary other wavelengths.

Structures like the one explained above can be envisaged to be written holographically, or using direct laser writing with subsequent infiltration of the structures in order to lower the refractive-index contrast.

### 4. Simulation Methods

Next, we choose a specific implementation of the presented model structure and calculate its reflectivity spectrum. For this structure we apply both the first-order approximation and numerical simulation using FIT. For the first-order calculation the equation for the backscattered power density of a structured film from reference [54] was used and simply divided by the intensity of the incident wave to obtain a reflectivity.

$$R = R_{\text{af}} + \frac{T_{\text{af}} k^2 \hat{\varepsilon}^2}{(4\pi)^2 l^2 \bar{\varepsilon}^2} \iint\limits_{\text{half of sphere } |\mathbf{k}_1|=k} |F_l(\mathbf{k}_1 - \mathbf{k}_0)|^2 \left[ T_{\text{fa}}^s (\mathbf{e}_0 \cdot \mathbf{e}_1^s)^2 + T_{\text{fa}}^p (\mathbf{e}_0 \cdot \mathbf{e}_1^p)^2 \right] d^2 k_1, \quad (2)$$

Here, $R$ and $T$ are the reflectivity and transmissivity of the film surface, respectively. The subscripts 'af' and 'fa' represent the transition from air into the film and vice-versa, respectively. $\mathbf{k}_0$ and $\mathbf{k}_1$ are the wave vectors of the incident and the reflected wave, respectively, and $k = |\mathbf{k}_0| = |\mathbf{k}_1|$. The side length of the structure is depicted as $l$ and $F_l$ is the spatial Fourier transform of the structure. Finally, the unit vector $\mathbf{e}_0$ is the polarization vector of the incident wave and $\mathbf{e}_1^s$ and $\mathbf{e}_1^p$ are the unit vectors for the $s$- and $p$-polarization of the scattered wave, respectively. Note, that the transmission of the scattered wave into air is polarization dependent, which is denoted by the superscripts $s$ and $p$ in the respective transmissivities. This equation can be connected to the Ewald sphere construction as following. The reflection is evaluated as a sum of the intensities scattered in the film into the backward direction and transmitted through the film-air interface. The integration on the half sphere in Eq. (2) represents the half of the Ewald sphere that contains only backward directions, which in our case is the upper half sphere representing scattering into the upper half space. The values of the Fourier transform of the structure $F_l$ on the surface of this sphere define the intensity of the scattered light for each direction of scattering. In our consideration $F_l$ has a form of a shell, which we call $k$-shell. In this case the values of $F_l$ on the Ewald sphere are non-zero only at their intersections (as shown in Fig 2). Additionally, Fresnel reflection at the film-air interface is taken into account for each polarization and direction. A more detailed explanation of this equation and its derivation are given in reference [54].

A volume of 6×6×40 µm³ was used. The lateral ($x$- and $y$-) dimensions (see Fig. 6) are big enough to be representative of a film-like structure with thickness of 40 µm. We assume that there is air above the film and below is a material that does not allow any reflection. The latter can either be an absorber or a lower half space filled with a material matched to the mean refractive index of the scattering structure. We use these asymmetric top and bottom boundary conditions to resemble a structural-color film as a coating on another material with similar refractive index as a typical arrangement for a structural color.

As a reference for the scattering behavior predicted with the first-order approximation we performed FIT simulations of the same system using the time domain solver of CST Microwave Studio [55]. The structure is excited by a plane wave that is normally incident and polarized in $x$-direction (coordinates defined as in Fig. 6). The lateral boundary conditions are defined as perfect electric and magnetic conductors in a way such that the incident wave is not disturbed. By this, the boundaries act as perfect mirrors and thus again emulate the behavior of an infinitely extending film. The error induced by using mirror boundary conditions was found to be negligible for the simulated width of 6 µm as no major changes in the spectrum occurred after changing the width. Below the structure there is an additional block of material with a refractive index matched to the mean refractive index of the structure to minimize reflection and ensure a controlled transition to the bottom boundary of the simulation volume. For the top and bottom of the simulation volume open boundaries were used which act as reflection-free terminations. The time-resolved reflected and transmitted fields are monitored in plane areas close to the upper and lower boundary of the simulation volume, respectively. The time signals are then converted into the frequency domain using a discrete Fourier transform. From the fields in the frequency domain the powers can be calculated. Finally, by normalization to the intensity of the incident wave and integration over the respective monitor planes the reflectivity and transmissivity are found. Using the spatial Fourier transform of the field distribution in a plane above the simulated structure we can also obtain the 2D distribution of the in-plane components of the scattered $k$-vectors which directly correspond to the angle distribution of scattered waves for the specified wavelength.

We want to emphasize that the reflectivity spectra shown in this paper do not exactly represent the color expected from a realistic structural color. We acquired the reflectivity by integrating over the whole upper half space whereas in realistic scenarios the recipient typically receives light reflected into very small solid angles, only. Furthermore, all simulations were considering normal incidence whereas in practical applications colors are often subject to diffuse illumination.

## 5. Results and Discussion

Figure 6 compares the reflectivity spectra of the idealized model structure as predicted by FIT and by first-order approximation without and with Fresnel reflection. The FIT simulation is represented by the black curve in Fig. 6 and is discussed first. It shows a strong reflectivity at wavelengths between approximately 620 and 665 nm. The steep edge on the long-wavelength side of this peak fits to the expected cut-off of the overlap between Ewald's sphere and the Fourier transform of the structure, leading to all wavelengths longer than 665 nm being transmitted. The residual reflectivity for these wavelengths is only due to Fresnel reflection at the interface, which for air and the used material with mean refractive index of 1.5 has a value of 0.04. The steep slope at the short wavelength side of this peak is defined by the angle of total internal reflection. Scattered light of wavelengths shorter than 617 nm is impinging on the film interface from inside at angles larger than the critical angle, thus is reflected back into the film via total internal reflection (also schematically seen as the orange arrows in Fig. 3).

Using the Ewald sphere construction, the scattering direction for incident light at a free-space wavelength of about 467 nm turns out to be perpendicular to the normal direction of incidence (also see Fig. 2(b)). As discussed in the theory part this is just the designed peak wavelength of 660 nm divided by $2^{1/2}$. At this wavelength we leave the backscattering regime as all shorter wavelengths are scattered forward. The blue curve in Fig. 6, which is discussed next, shows the backscattering efficiency inside the material without consideration of any reflections at interfaces (Fresnel reflection as well as total internal reflection). In this case, the short-wavelength cut-off at around 467 nm is caused by the above-mentioned transition from backward to forward scattering, only. The long-wavelength edge is found at approximately 660 nm as expected. In the regime between the two cut-offs the reflectivity increases with decreasing wavelength.

When we take into account reflections from interfaces (Fresnel reflection as well as total internal reflection) the red curve in Fig. 6 results. For normal incidence and given the effective refractive index of $\bar{n} = 1.5$, the red curve shows that total internal reflection of the scattered wave sets in at a wavelength of about 617 nm. This finding corresponds very well with the decrease of the reflectivity at the short-wavelength side of the peak from FIT simulation (black curve). Thus, the calculated peak in the red curve matches the peak in the simulated black curve. This confirms the validity of our approach and shows that total internal reflection may play a significant role for the selectivity of structural-color films.

At shorter wavelengths we observe deviations between the calculated red curve and the simulated black curve. As discussed before these are due to multiple scattering which occurs for shorter wavelengths and, naturally, is not included in our first-order calculation [17], however, shows up in the brute force FIT-simulation. Specifically, there is a peak at wavelengths around 467 nm which is likely due to light scattered twice. The high relative strength of this secondary scattering can be explained by the film geometry. Light scattered at 90 degrees, due to the infinite horizontal propagation path inside the film, has a probability of one to scatter again, with 50 percent of that second order scattered light going upwards, thus partly coupling out and adding to the reflected signal, and 50 percent going downwards into the forward direction, either being absorbed, or transported away reflection free, eventually. Wavelengths shorter than 467 nm are increasingly scattered into the forward direction which goes along with limiting of the propagation path, again, thus reducing the probability for a second scattering event before the light can exit the film in the forward direction. We further obtain the cut-off wavelengths on both sides of this broad peak by consecutively using the Ewald sphere construction twice. The total internal reflection preventing the coupling out of doubly scattered light in the backward direction (reflection regime) sets in at 374 nm for light that was initially scattered into a forward direction and at 544 nm wavelength for light that was initially scattered into a backward direction. These two wavelengths match the limits of the secondary scattering peak seen in the simulated reflectivity spectrum (black curve).

Whether the mechanisms described here might be responsible for the double-peaked reflectivity spectra often observed in bird feathers (e.g. [13]) remains to be clarified in future research. The main difference between the structures considered here and biological ones is the refractive-index contrast. The biological structures are usually binary structures from keratin/chitin filled with air. The refractive-index contrast of 1 to approximately 1.55 [56] would lead to a stronger contribution of higher-order scattering than what is observed in our structures. The two peaks would then almost merge and numerical wave propagation simulations would be required to get the exact reflectivity spectra. Still, the geometrical Ewald sphere construction will allow predicting important scattering features such as the long-wavelength reflectivity edge.

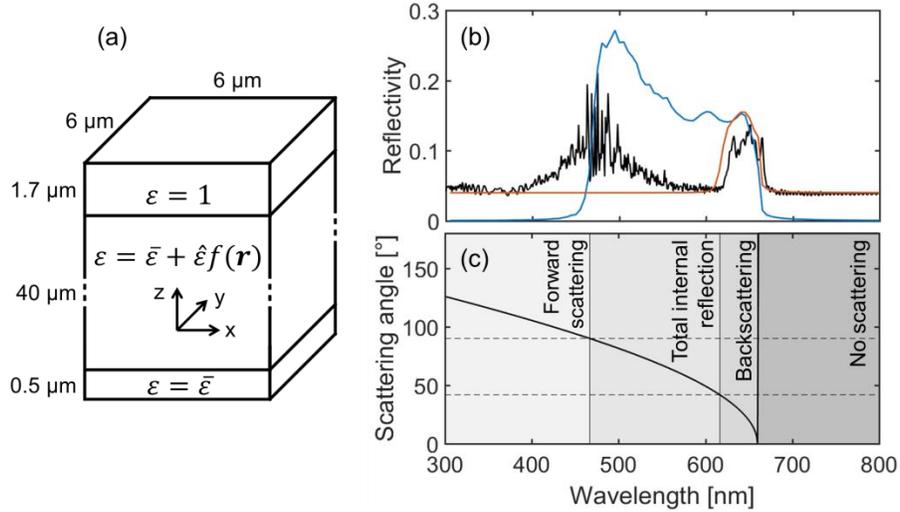

Fig. 6. (a) Schematic of the simulated volume. (b) FIT-simulated reflectivity spectrum of the model structure with 40 μm thickness, a mean refractive index of 1.5 and a refractive-index standard deviation of $\sigma = 0.025$ (black curve). In comparison to that, the first-order approximation calculated without (blue curve) and with (red curve) interface reflections (Fresnel reflection as well as total internal reflection) taken into account. The incident wave impinges normally onto the film from the top. (c) Scattering angle for first-order scattering with respect to the normal for a scattering structure that exhibits a perfect spherical shell in $k$-space with radius $K = 0.0286 \text{ nm}^{-1}$. The horizontal, dashed lines mark the 90 degree angle and the angle of total internal reflection at the film-air interface.

Next, we investigate the use of the Ewald sphere construction in predicting the directionality of the scattered light by comparing it to the scattering directions above the film calculated using FIT in dependence on the wavelength. In Fig. 7 we illustrate the directions of scattered light for four different wavelengths (a-d). The plots are in the $(k_x, k_y)$-plane of the reciprocal space as indicated by the arrows in the rightmost graph (Fig. 7(d)). The Fourier transform decomposes the fields into a sum of plane waves. Each point in the diagram corresponds to a plane wave with a direction of propagation given by the respective $k_x$ and $k_y$ while $k_z$ is implicitly given by $k_{\text{air}}^2 = k_x^2 + k_y^2 + k_z^2$, where $k_{\text{air}}$ is the wave number of the incident and reflected waves in air, $k_{\text{air}} = |\mathbf{k}_0|/\bar{n} = |\mathbf{k}_1|/\bar{n}$. The color at each $(k_x, k_y)$-position is a measure of the intensity scattered into the corresponding direction. Red colors indicate large values and dark blue means zero. All plotted points are within the white circle, the radius of which equals the length $k_{\text{air}}$ of the wave vector in air. Points on this circle would indicate propagation in directions parallel to the film surface, i.e. $k_{\text{air}}^2 = k_x^2 + k_y^2$ and $k_z = 0$. On the other hand, the center of the circle stands for propagation normal to the film surface, i.e. $k_{\text{air}} = k_z$ and $k_x = k_y = 0$. It should be mentioned that the graphs are normalized to the length of the wave vector such that the radius of the white circle is the same for all wavelengths. Note further that the scattering angles in air and inside the film material are related to each other via Snell's law. With respect to the normal, light scattered inside the structure into the angular range between zero degree and the angle of total internal reflection is refracted into the full angular range between zero and 90 degrees in air.

From the Ewald sphere construction a diffusive distribution of *k*-vectors is expected in the secondary scattering peak because after two scattering events there is hardly a preferential direction anymore [17]. The simulated *k*-vector distribution at a wavelength of 465 nm is shown in Fig. 7(a). The slight anisotropy in the distribution can be attributed to polarized excitation and polarization dependent Fresnel reflection. Towards longer wavelengths single scattering predominates and the scattered wave vectors are expected to lie on the surface of a gradually closing cone. Figure 7(b) shows the *k*-vector distribution at 645 nm where a clear ring is seen. This means that light is mainly scattered for a certain $k_r^2 = k_x^2 + k_y^2$, i.e. a certain $k_z$, which corresponds to a scattering angle $\alpha$ with respect to the plane normal, $k_r = k_0 \sin(\alpha)$. At 660 nm or slightly above where the Ewald sphere just touches the *k*-shell a filled spot in the wave vector distribution is observed, corresponding to a narrower but fully filled cone of scattering angles (Fig. 7(c) at 665 nm). For even longer wavelengths neither backward nor forward scattering occurs anymore and light is simply transmitted because the Ewald sphere does no longer intersect with the *k*-shell anymore due to its decreased radius. The residual reflectivity is only caused by the weak Fresnel reflection from the film surface in normal backward direction which corresponds to a point at zero in the ($k_x$,$k_y$)-plane (Fig. 7(d) at 750 nm). Thus, the Ewald sphere may serve as a tool for predicting several important aspects of scattering in structural colors such as the spectral position of the reflectivity peak and the directions of scattering.

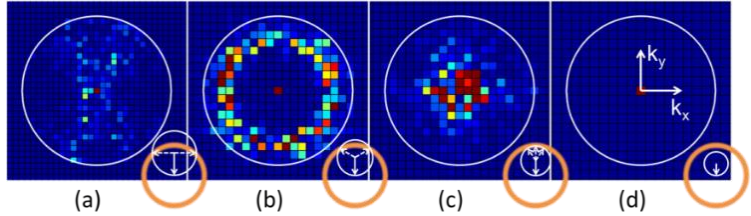

Fig. 7. Absolute value squared of the Fourier transform of the scattered fields at certain wavelengths calculated in an (*x*,*y*)-plane above the structure and encoded in color; red corresponds to large values and dark blue to zero. We obtained the squared amplitude by Fourier transforming the fields $E_x$, $E_y$, $E_z$ and subsequently calculating $A^2(k_x, k_y) = |F\{E_x\}|^2 + |F\{E_y\}|^2 + |F\{E_z\}|^2$, where F denotes the two-dimensional Fourier transform in the (*x*,*y*)-plane. *k*-vectors parallel to the film surface mark the maximum spatial frequency which is indicated by the white circle. The red point in the center represents the incident and the Fresnel-reflected wave which are propagating normally to the film interface and therefore correspond to a spatial frequency of zero in the ($k_x$,$k_y$)-plane. The shown wavelengths are (a) 465 nm, (b) 645 nm, (c) 665 nm and (d) 750 nm, selected to lie, respectively, in the second-order scattering peak, in the conical reflection regime, at the reflectivity edge, and in the scattering-free spectral regime (cf. Fig. 6(b)). At the bottom right for each case the corresponding Ewald sphere construction is shown schematically.

With these results we have shown that the first-order approximation can be used for weakly scattering structures with a low refractive-index contrast. For structures with higher contrast the first-order approximation will not yield quantitatively correct results. However, the directions of scattering in dependence on the wavelength can still roughly be estimated by the Ewald sphere construction. For example, we used this for the optimization of artificial structural colors based on photonic glasses [57]. Considering this option we believe that the approach presented here can be generally useful for the understanding of light propagation in disordered structures with partial order.

## 6. Conclusions

We have shown the applicability of the first-order Born approximation and the Ewald sphere construction to effectively describe scattering in disordered structures used to realize structural colors. We have applied them to predict several features in the scattering of light normally incident on a structural-color film, such as a sharp reflectivity edge towards long wavelengths, an opening scattering angle towards shorter wavelengths and finally the transition into multiple scattering. We further showed that total internal reflection of scattered light at the film surface leads to a suppression of backscattered light even for wavelengths where the scattering as such would still admit backscattering. This effect leads to the formation of a narrow spectral peak with a sharp edge on the short-wavelength side and thus increases the spectral selectivity of the reflection. The vivid geometrical representation of the Ewald sphere and its simple numerical applicability make the Ewald sphere construction an excellent tool to study structural colors and other scattering problems.

As stated before, we only consider normal incidence of light on the scattering film but in realistic situations the illumination is often diffuse. While the full numerical simulation of such a scenario poses quite a challenge it should be much simpler to add up the reflectivity predicted by the first-order approximation for several incidence directions to get a good approximation of the total scattering. This is, however, beyond the scope of this paper and remains to be investigated.

Another thing to note is that the presented approach works best for weak overall scattering. However, for a thick disordered structure with weak local scattering a combined approach where the Ewald sphere construction is used to predict the angular distribution of each scattering event could be easily implemented in a Monte Carlo simulation. Such an approach could also be used for structures producing strong reflectivity contrasts and might be an interesting next step in the modeling of disordered structures for structural color, optical diffusion and random lasing.

## Funding

This work was supported by the German Research Foundation (DFG) via priority program SPP 1839 "Tailored Disorder" and via the Collaborative Research Centre SFB 986 "Tailor-Made Multi-Scale Materials Systems". Funded by the Deutsche Forschungsgemeinschaft (DFG, German Research Foundation) – Projektnummer 392323616 and the Hamburg University of Technology (TUHH) in the funding program "Open Access Publishing".

## Acknowledgements

We thank CST, Darmstadt, Germany for providing us their Microwave Studio software.